\documentclass[aps,prb,reprint,twocolumn,superscriptaddress,showpacs,floatfix,longbibliography]{revtex4-1}

\usepackage{amsmath}
\usepackage{color}
\usepackage{mathrsfs}
\usepackage{dcolumn}
\usepackage{bm}
\usepackage{multirow}
\usepackage{graphicx}
\usepackage{rotating}
\usepackage{layout}
\usepackage[version=3]{mhchem}
\usepackage{braket}
\usepackage{setspace}
\usepackage{float}
\usepackage{graphicx}
\usepackage{subcaption}
\usepackage{verbatim}
\usepackage{siunitx}
\usepackage{xcolor}
\usepackage{caption}
\allowdisplaybreaks
\usepackage{xfrac}

\usepackage[colorlinks,citecolor=blue,urlcolor=blue,bookmarks=false,hypertexnames=true]{hyperref}

\begin{document}
\title{Reducing quantum resources for ADAPT-VQE via plateau-operator elimination and correlated mean-field downfolding}
\date{\today}
\author{Phuoc Minh Vo}
\affiliation{Faculty of Physics and Engineering Physics, University of Science, Ho Chi Minh City 70000, Vietnam}
\affiliation{Vietnam National University, Ho Chi Minh City 70000, Vietnam}
\author{Thai Cong Ngoc Vu}
%\email{thai.vucongngoc@phenikaa-uni.edu.vn}
\affiliation{Phenikaa Institute of Advanced Study (PIAS), Phenikaa University, Hanoi 12116, Vietnam.}
\author{Thien Ngoc Tran}
\affiliation{Faculty of Physics and Engineering Physics, University of Science, Ho Chi Minh City 70000, Vietnam}
\affiliation{Vietnam National University, Ho Chi Minh City 70000, Vietnam}
\author{Hoang Thanh Nguyen}
\affiliation{Institute of Advanced Technology, Vietnam Academy of Science and Technology, Ho Chi Minh City 70 000, Vietnam}
\author{Lan Nguyen Tran}
\email{tnlan@hcmus.edu.vn}
\affiliation{Faculty of Physics and Engineering Physics, University of Science, Ho Chi Minh City 70000, Vietnam}
\affiliation{Vietnam National University, Ho Chi Minh City 70000, Vietnam}
\date{\today}

\date{\today}

\begin{abstract}
Adaptive Derivative-Assembled Problem-Tailored variational quantum eigensolvers (ADAPT-VQE) represent one of the most promising approaches for quantum chemistry on near-term quantum devices. However, their optimization is slow and may stall due to vanishing parameters and redundant operators in the ansatz. In this work, we propose a simple strategy of operator elimination that removes non-contributing operators from the pool once they are detected, enabling the optimization to continue progressing toward convergence. We examine two variants, with and without pool restoration after elimination, and find that the former converges more smoothly and faster than the latter and the standard ADAPT-VQE. To capture dynamical correlations between the active space and its environment, we combine ADAPT-VQE with our recently developed downfolding approach, the one-body downfolding framework (OBDF). In OBDF, the bare molecular Hamiltonian in the active space is replaced by a correlated effective Hamiltonian that incorporates dynamical correlation effects outside the active space. We benchmark our implementation on a linear \ce{H_6} chain, an \ce{H_6} lattice, an \ce{H_6} ring, and the \ce{N_2} molecule using the OpenFermion simulator. Our results show that operator elimination significantly reduces circuit depth and iteration count, and that OBDF-ADAPT-VQE yields energies closer to the full configuration interaction (FCI) reference than the standard approach within the same active space.

\end{abstract}

\maketitle

\section{Introduction}
Accurate computation of the electronic structure of molecules and materials is a central problem in quantum chemistry and condensed matter physics. The exact solution of the many-body Schr\"{o}dinger equation via full configuration interaction (FCI) is formally exact but scales exponentially with the number of electrons, making it tractable only for very small systems~\cite{helgaker2000}. Approximate methods, such as coupled cluster~\cite{bartlett2007} and perturbation theory~\cite{MP2}, achieve high accuracy for weakly correlated systems but break down in the presence of strong correlation and similarly suffer from steep computational scaling. The prospect of using quantum computers to simulate quantum systems efficiently, first envisioned by Feynman~\cite{feynman1982} and later formalized by Lloyd~\cite{lloyd1996}, has therefore attracted great attention as a means of circumventing these classical bottlenecks. In the current noisy intermediate-scale quantum (NISQ) era~\cite{preskill2018}, quantum devices with tens to hundreds of qubits are limited by noise, decoherence, and restricted circuit depth, motivating the development of hybrid classical--quantum algorithms to minimize quantum resource requirements~\cite{cao2019,mcardle2020}.

The variational quantum eigensolver (VQE), developed by Peruzzo \textit{et al.}~\cite{peruzzo2014}, is one of the most widely used algorithms for estimating the ground-state energies of molecular systems. VQE operates by preparing a parameterized trial state (ansatz) on a quantum processor and minimizing its energy expectation value using a classical optimizer in a self-consistent loop. This hybrid architecture keeps the quantum circuit depth relatively shallow while delegating the optimization to classical hardware. Subsequently, McClean \textit{et al.}~\cite{mcclean2016} provided a theoretical framework for VQE and demonstrated its applicability to molecular simulation. The unitary coupled cluster singles and doubles (UCCSD) ansatz has emerged as a chemistry-inspired ansatz for VQE, offering a systematically improvable approximation to the wavefunction~\cite{peruzzo2014,shen2017,romero2018}. However, standard VQE with a fixed ansatz faces several drawbacks. The choice of ansatz is non-trivial, and a poorly chosen ansatz may fail to capture the essential physics of the system, leading to poor convergence or large errors~\cite{fedorov2022vqe}. Furthermore, McClean \textit{et al.}~\cite{mcclean2018} demonstrated the existence of so-called \textit{barren plateaus} in the optimization landscape, where gradients of the cost function vanish exponentially with increasing system size, rendering classical optimization infeasible. In addition, the UCCSD circuit depth scales as $\mathcal{O}(N^5)$ with the number of spin-orbitals (qubits) $N$, which quickly becomes prohibitive for NISQ devices~\cite{romero2018}. These limitations motivate the search for more resource-efficient and adaptive quantum algorithms.

Adaptive Derivative-Assembled Problem-Tailored VQE (ADAPT-VQE), introduced by Grimsley \textit{et al.}~\cite{grimsley2019}, tackles the fixed-ansatz problem by building the wavefunction incrementally. Rather than committing to a predetermined circuit, the ansatz grows one operator at a time. At each step, the operator with the largest energy gradient magnitude is selected from a predefined pool and appended to the circuit. For small molecules, this produces compact, system-tailored circuits that reach chemical accuracy with far fewer parameters than fixed-ansatz VQE~\cite{grimsley2019,Tilly2022}. The approach has since been extended to qubit-based operator pools~\cite{tang2021,Anastasiou2024Tetris,Feniou2023}, spin-complemented pools~\cite{shkolnikov2023}, and noise-aware variants~\cite{gomes2021,feniou2025greedy}.

However, the optimization in ADAPT-VQE is sensitive to both the reference state~\cite{VanDyke2024} and the operator pool. When the Hartree--Fock (HF) reference is qualitatively incorrect, convergence can be slow~\cite{Grimsley2023}. Another issue is that the ADAPT-VQE ansatz often consumes a large number of CNOT layers contributed by redundant operators~\cite{rossi2026resource}, hampering the practical scalability of the algorithm~\cite{tancara2026}. Several methods have recently been developed to mitigate this issue, including Pruned-ADAPT-VQE, proposed by Vaquero-Sabater and co-workers~\cite{prunedADAPT}, and Param-ADAPT-VQE, developed by He and co-workers~\cite{compactADAPT}. In Pruned-ADAPT-VQE, each operator is assigned a score based on its parameter value and its position in the ansatz after optimization, and operators with low scores are subsequently discarded. A dynamic threshold based on the parameters of recent operators is also applied to enable efficient convergence. Param-ADAPT-VQE, on the other hand, selects excitation operators based on a parameter-based criterion rather than the traditional gradient-based metric.

An alternative approach for reducing quantum resource requirements is Hamiltonian downfolding, in which the full-space Hamiltonian is mapped to an effective Hamiltonian defined for an active space, with the effects of the external (inactive) space folded in through renormalization. Double CC (DCC) downfolding, developed by Bauman, Kowalski, and coworkers~\cite{bauman2019,kowalski2020,kowalski2021}, goes beyond the bare active space by constructing an effective Hamiltonian that explicitly incorporates correlation effects from the outside into the active space. This effective Hamiltonian acts on a reduced orbital space and yet encodes information about the full many-body problem, thereby allowing accurate quantum simulations to be carried out with fewer qubits and shallower circuits. The downfolding formalism has been demonstrated to be a powerful preprocessing method for reducing the cost of quantum simulations~\cite{bauman2019,kowalski2020}, and its combination with VQE-type algorithms has been explored~\cite{bauman2022}. In the same vein, we have recently developed one-body downfolding (OBDF) in a resource-efficient manner~\cite{OBMP2-JPCA2023}. In the OBDF framework based on one-body M{\o}ller--Plesset second-order perturbation theory (OBMP2)~\cite{OBMP2-JCP2013,OBMP2-JPCA2021,OBMP2-PCCP2022,OBMP2-JPCA2024,OBMP2-JCP2025}, the effect of external (inactive) orbitals is incorporated into the active-space Hamiltonian through a correlated one-body operator that modifies only the one-body part of the Hamiltonian. The resulting effective Hamiltonian retains the same operator structure as the bare active-space Hamiltonian and therefore requires no additional quantum resources beyond those needed for the bare problem.

In the current work, to address the redundant operator problem in ADAPT-VQE, we propose a simple and direct elimination scheme that removes operators with vanishing parameters, allowing the optimization to continue progressing without unnecessary operators. The convergence is considerably accelerated, and the resulting energies remain close to or even lower than those of standard ADAPT-VQE. We assess two elimination variants: one with pool restoration and one without. We find that the operator elimination scheme without pool restoration converges more smoothly and faster than the one with pool restoration. We show that, although our direct elimination method is less sophisticated than Pruned-ADAPT-VQE and Param-ADAPT-VQE, it yields promising results and may be useful for practical enhancement. Furthermore, to capture correlations between the active space and the external environment, we combine ADAPT-VQE with OBDF, in which the bare molecular Hamiltonian is replaced by a correlated effective Hamiltonian that incorporates dynamical correlation outside the active space. All ADAPT-VQE calculations were performed using the OpenFermion simulator~\cite{liu2021efficient}. We benchmark our approaches on a set of representative systems, including linear \ce{H_6} chains, \ce{H_6} lattices, \ce{H_6} rings, and the \ce{N_2} molecule across various geometries. Our results demonstrate that ADAPT-VQE with operator elimination converges significantly faster and requires fewer circuit layers than the standard approach, while ADAPT-VQE combined with OBDF consistently yields energies closer to the FCI reference than conventional ADAPT-VQE within the same active space, highlighting the benefit of incorporating external correlation through downfolding.

\begin{figure*}[t!]
\includegraphics[width=1.0\linewidth]{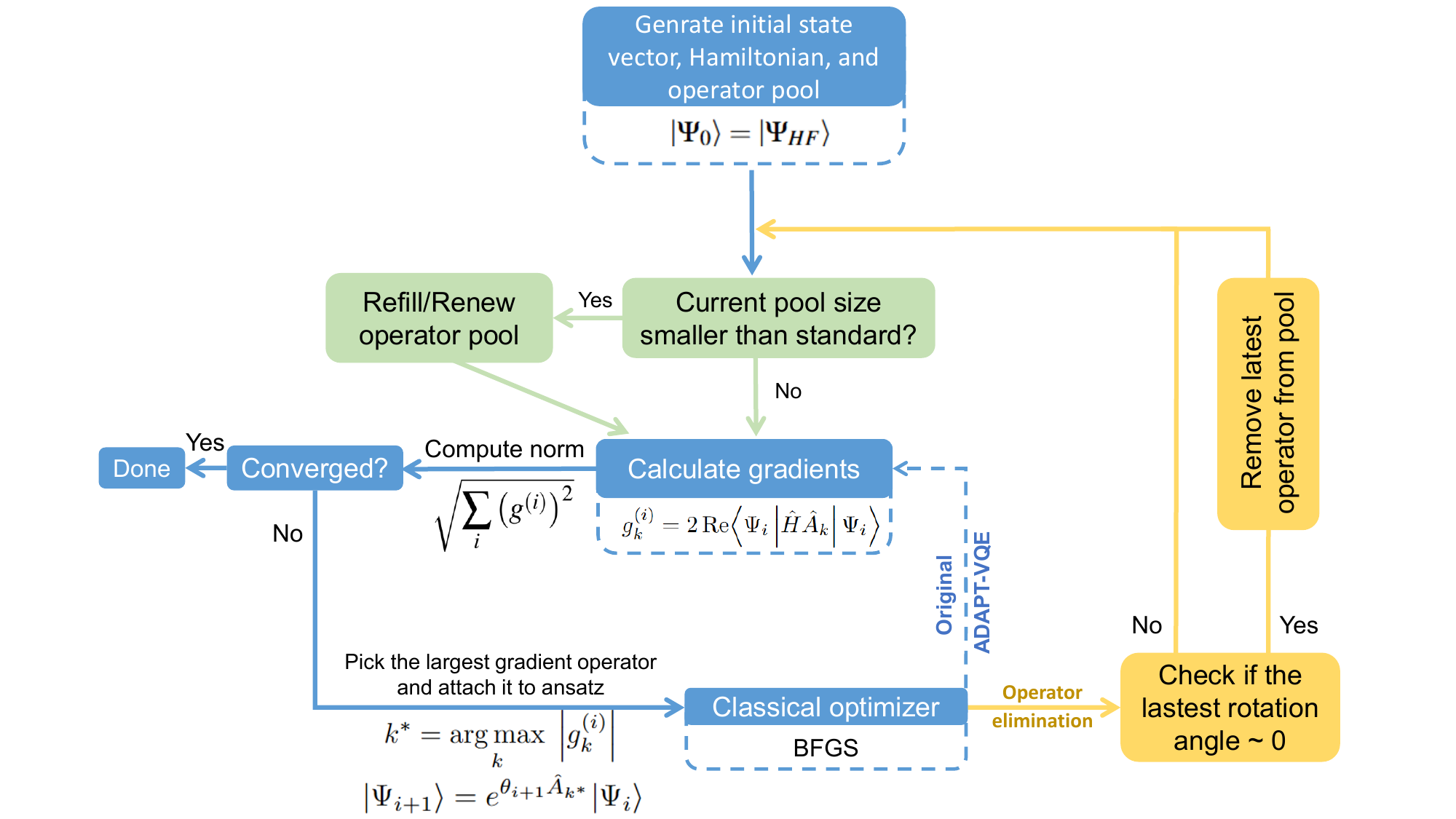}
    \caption{The plateau-operator elimination is incorporated into the original ADAPT-VQE workflow. After the classical optimization, the elimination scheme is applied to remove operators with vanishing parameters.}
    \label{fig:workflow}
\end{figure*}

\section{Theory}{\label{sec:theo}}
\subsection{ADAPT-VQE overview}

VQE relies on the variational principle asserting that the ground-state energy $E_{0}$ is bounded from above by the expectation value of the Hamiltonian
\begin{align}
    E_0 \leq \frac{\left<\psi\right| \hat{H} \left|\psi\right>}{\left<\psi|\psi\right>},
\end{align}
where the Hamiltonian $\hat{H}$ is expressed as:
\begin{align}
  \hat{H} = \sum_{pq}h^{p}_{q} \hat{a}_{p}^{q} + \tfrac{1}{2}\sum_{pqrs}g^{p r}_{q s}\hat{a}_{p r}^{q s}\label{eq:h1}
\end{align}
with the indices $\left\{p, q, r, s\right\}$ running over all general spin orbitals. The excitation operators appearing in Eq.~\ref{eq:h1} are defined as $\hat{a}_{p}^{q}=\hat{a}^{\dagger}_p \hat{a}_q$ and $\hat{a}_{p r}^{q s} = \hat{a}^{\dagger}_p \hat{a}^{\dagger}_r \hat{a}_s \hat{a}_q$, respectively; and $h_{pq}$ and $ g_{pq}^{rs}$ are one- and two-electron integrals. 

For a given Hamiltonian, the objective of VQE is to minimize its expectation value \cite{peruzzo2014}. Assuming the trial wavefunction $\left|\psi\right>$ is normalized, the VQE optimization problem is formally expressed as:
\begin{align}
    E_{\text{VQE}} = \min_{\boldsymbol \theta} \left<\boldsymbol \psi \right| U^\dagger(\boldsymbol \theta) \hat{H} U(\boldsymbol \theta) \left|\boldsymbol \psi \right> \label{eq:E-vqe}
\end{align}
The central limitation of VQE lies in the manual selection of the ansatz. A poorly chosen form not only leads to the barren plateau problem\cite{mcclean2018} but also incurs substantial computational overhead in searching for a suitable circuit structure. ADAPT-VQE, introduced by Grimsley et al. \cite{grimsley2019}, addresses this limitation by constructing the ansatz iteratively, selecting operators based on information from the energy gradient.

The ADAPT-VQE algorithm begins from a reference state $|\Psi_0\rangle = |\Psi_\mathrm{HF}\rangle$ and maintains an operator pool $\{\hat{A}_k\}$ of anti-Hermitian single and double fermionic excitation operators. At each iteration $i$, the energy gradient with respect to each pool operator is evaluated as
\begin{equation}
    g_i^{(k)} = 2\,\mathrm{Re}\!\left\langle \Psi_{i-1} \left| \hat{H}\hat{A}_k 
    \right| \Psi_{i-1} \right\rangle.
\end{equation}
The operator carrying the largest gradient magnitude,
\begin{equation}
    k^* = \underset{k}{\mathrm{arg\,max}}\; \left|g_i^{(k)}\right|,
\end{equation}
is appended to the ansatz, giving the updated state
\begin{equation}
    |\Psi_{i+1}\rangle = e^{\theta_{i+1}\hat{A}_{k^*}}|\Psi_i\rangle.
\end{equation}
All variational parameters $\{\theta_1, \ldots, \theta_{i+1}\}$ are then re-optimized jointly using a classical optimizer, such as BFGS or gradient descent, to minimize the energy expectation value
\begin{equation}
    E = \left\langle \Psi_{i+1}\right| \hat{H} \left|\Psi_{i+1}\right\rangle.
\end{equation}
The procedure is iterated until the pool gradient norm drops below a prescribed convergence threshold $\epsilon$,
\begin{equation}
\sqrt{\sum_{k=1}^L\left(\mathbf{g}^{(k)}\right)^2} < \epsilon,
\end{equation}
at which point the ansatz is considered converged. Here, $L$ is the number of operators in the pool. As a result, the circuit grows incrementally, adding only the operator that contributes most to the energy reduction at each step. In contrast to fixed-structure ansätze, such as UCCSD\cite{peruzzo2014, shen2017, romero2018}, this avoids including operators that are energetically irrelevant.

\begin{figure*}[t!]
    \centering    \includegraphics[width=0.8\linewidth]{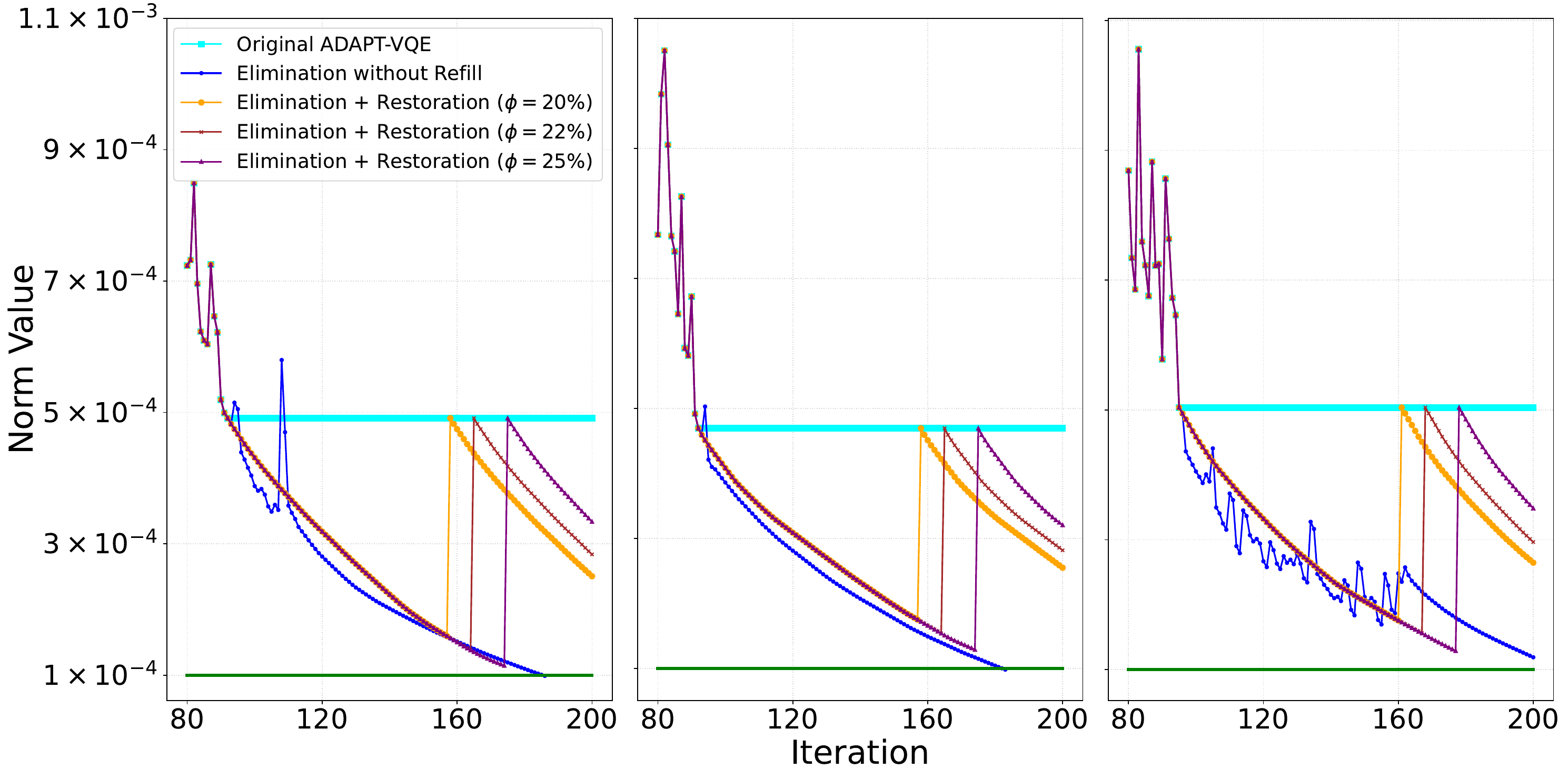}
    \caption{Gradient norm convergence for the hydrogen chain at $r = 2.25$~\unit{\angstrom} (left), $2.55$~\unit{\angstrom} (middle), and $2.85$~\unit{\angstrom} (right). There are three thresholds of elimination, at which the pool is restored: $\phi = 20\%$  (orange), $22\%$ (purple), and $25\%$ (pink). The horizontal green line indicates the convergence threshold $\epsilon = 10^{-4}$.}
    \label{fig:H6Chain_norm_conv}
\end{figure*}

\subsection{Eliminating plateau operators} 
\label{subsec:epo}

One of the main issues in gradient-based optimization is the emergence of \textit{barren plateaus}, regions where energy gradients vanish exponentially with system size~\cite{mcclean2018}. In ADAPT-VQE, this happens when certain pool operators $\hat{A}_k$ yield vanishingly small parameters. We classify an operator $\hat{A}_k$ as a plateau operator at the $i$-th iteration if its corresponding rotation angle magnitude $|\theta_i|$ falls below a prescribed threshold $\delta$,
\begin{equation}
    |\theta_i| < \delta,
    \label{eq:plateau_criterion}
\end{equation}
where $\delta$ is very small (nearly zero). 

If the plateau condition is met, that operator is eliminated from the pool. This is not because such operators are physically irrelevant, but because the current ansatz $|\Psi_i\rangle$ does not have sufficient overlap with the subspace coupled by $\hat{A}_k$. Such operators have a negligible contribution to the energy reduction at the current iteration, leading to unnecessary gate overhead if selected in subsequent steps, leading to computational inefficiency and optimization instability.
\begin{comment}
\begin{equation}
    g^{(k)}_i  \approx 0.
    \label{eq:small_grad}
\end{equation}
\end{comment}

The overall scheme of plateau-operator elimination wrapped around the original ADAPT-VQE is presented in Figure~\ref{fig:workflow}. To reduce the measurement overhead associated with plateau operators, the pool $\mathcal{P}$ is pruned at each iteration. All selected operators satisfying Eq.~\eqref{eq:plateau_criterion} are excluded from the selection step, and operator growth is restricted to the active subset,
\begin{equation}
    \mathcal{P}^{(i)}_{\mathrm{active}} = 
    \left\{ \hat{A}_k \in \mathcal{P} \;\middle|\; 
    \left|\theta_i^{(k)}\right| \geq \delta \right\},
    \label{eq:active_pool}
\end{equation}
reducing the number of gradient evaluations per iteration from $|\mathcal{P}|$ to $|\mathcal{P}^{(i)}_{\mathrm{active}}|$ and directly lowering the quantum circuit measurement cost. An operator classified as a plateau operator at iteration $i$ may develop a non-negligible gradient at a later iteration $j > i$. We therefore restore the pruned pool $\mathcal{P}^{(i)}_{\mathrm{active}}$ to its initial size $|\mathcal{P}|$ when the ratio of eliminated operators exceeds a limiting threshold $\phi$.

The elimination of plateau operators is beneficial in two ways. First, it accelerates convergence by focusing on operators that provide the largest energy reduction. Second, it lowers circuit depth by avoiding low-gradient operators that would otherwise increase circuit complexity without meaningfully lowering the energy.

\begin{figure*}[t!]
    \centering
\includegraphics[width=0.8\linewidth]{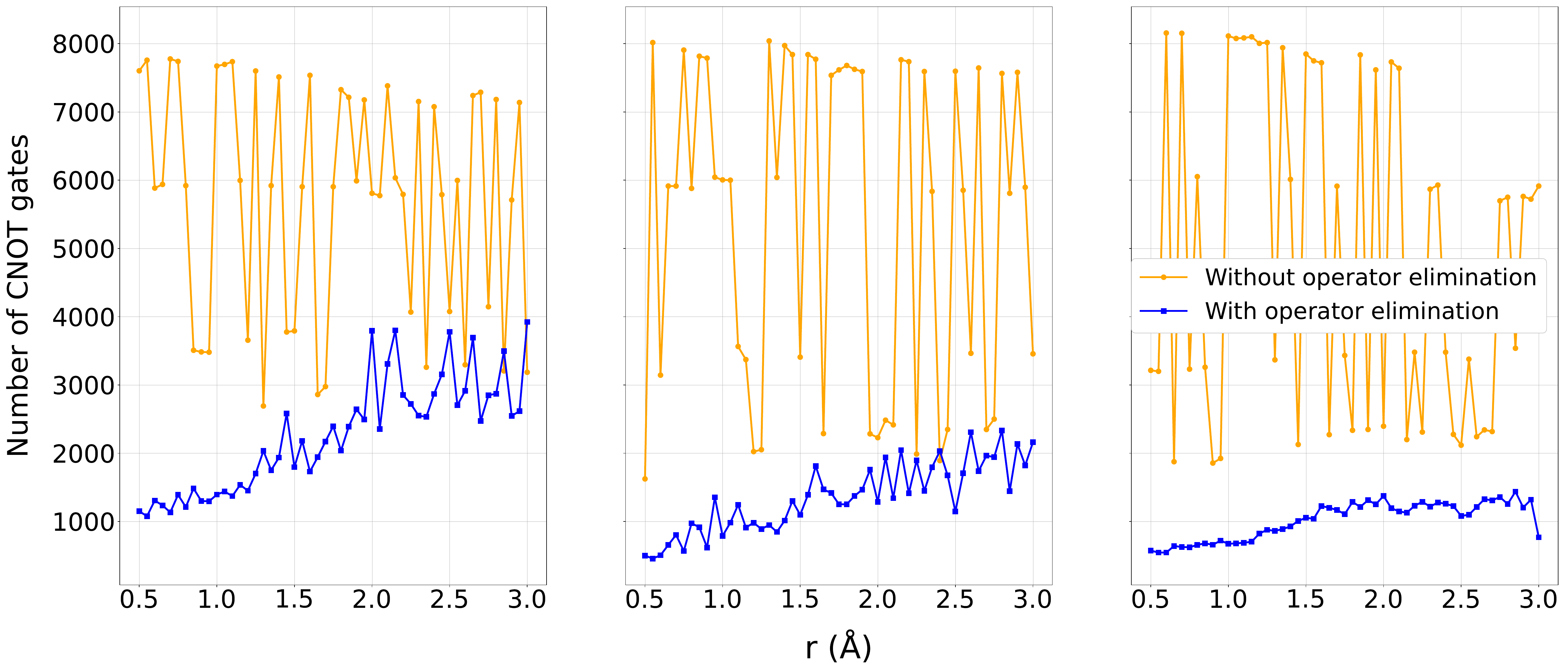}
    \caption{CNOT gate counts for ADAPT-VQE with and without operator elimination at different bond lengths of H$_6$ chain, H$_6$ ring, and H$_6$ lattice.}
    \label{fig:cnot_h6systems}
\end{figure*}

\subsection{ADAPT-VQE in the OBDF framework}

The OBMP2 approach~\cite{OBMP2-JCP2013,OBMP2-JPCA2021} is derived from the canonical 
transformation~\cite{CT-JCP2006,CT-JCP2007,CT-ACP2007,CT-JCP2009,CT-JCP2010,
CT-IRPC2010}, in which the molecular Hamiltonian $\hat{H}$ is similarity-transformed 
by a unitary operator $e^{\hat{A}}$:
\begin{equation}
    \hat{\bar{H}} = e^{\hat{A}^\dagger} \hat{H} e^{\hat{A}},
\end{equation}
where $\hat{A} = \hat{A}_\mathrm{D}$ is anti-Hermitian and carries MP2 amplitudes \cite{OBMP2-JCP2013}. Applying the Baker--Campbell--Hausdorff (BCH) expansion and retaining only one-body 
contributions via the cumulant 
approximation~\cite{cumulant-JCP1997,cumulant-PRA1998,cumulant-CPL1998,
cumulant-JCP1999}, the resulting effective Hamiltonian takes the form
\begin{equation}
    \hat{H}_\mathrm{OBMP2} = \hat{\bar{F}} + \bar{C},
\end{equation}
with a correlated Fock matrix $\bar{f}^p_q = f^p_q + v^p_q$, where $v^p_q$ encodes 
the dynamic correlation correction. Diagonalizing $\bar{f}^p_q$ yields a set of 
correlated molecular orbitals that implicitly incorporate dynamic correlation at 
$\mathcal{O}(N^5)$ cost, analogous to standard MP2.

To construct an effective active-space Hamiltonian, we employ the double unitary coupled-cluster (DUCC) downfolding 
framework~\cite{bauman2019downfolding,kowalski2021dimensionality}. Restricting the external cluster operator to double excitations involving at least one inactive orbital, $\hat{A}_\mathrm{ext} = \hat{A}^D_\mathrm{ext}$, and applying the OBMP2 approximation to the BCH commutator terms, the downfolded Hamiltonian projected onto the active space 
yields the one-body downfolded (OBDF) active-space Hamiltonian
\begin{equation}
    \hat{H}_\mathrm{OBDF} = \hat{H}_\mathrm{CAS} + \hat{v}^\mathrm{ext}_\mathrm{OBMP2},
    \label{eq:obdf_ham}
\end{equation}
where $\hat{H}_\mathrm{CAS}$ is the bare active-space Hamiltonian and $\hat{v}^\mathrm{ext}_\mathrm{OBMP2}$ is a one-body external correlation potential. Since the downfolding correction is purely one-body, $\hat{H}_\mathrm{OBDF}$ retains 
the same two-body operator structure as $\hat{H}_\mathrm{CAS}$, requiring no additional quantum resources beyond those of the bare active-space problem.

Within this framework, ADAPT-VQE is employed to solve the eigenvalue problem defined by $\hat{H}_\mathrm{OBDF}$. The correlated molecular orbitals obtained from diagonalizing $\bar{f}^p_q$ serve as the one-body basis for the active space, ensuring that the adaptively grown ansatz operates on a basis that already encodes dynamic correlation. Consequently, the ADAPT-VQE procedure requires fewer ansatz layers to 
achieve a given accuracy compared to using bare Hartree--Fock orbitals. This combination of OBMP2 downfolding with ADAPT-VQE, hereafter referred to as OBDF-ADAPT-VQE, constitutes the central methodological contribution of the present work. All OBMP2 and OBDF calculations are carried out within a local version of 
PySCF~\cite{pyscf-2018}, and ADAPT-VQE is implemented using the Openfermion framework~\cite{liu2021efficient,Yordanov2021QEB}.

\section{Results and discussion}
{\label{sec:result}}

\subsection{Gradient norm convergence}
Let us first assess the convergence of the gradient norm. We employ two variants of the operator elimination scheme: (i) permanent elimination without refilling and (ii) elimination with pool restoration. In the second scheme, the pool is 
restored to its original composition once the fraction of eliminated operators reaches a cutoff threshold ($\phi = 20\%$, $22\%$, or $25\%$).

The \ce{H6} chain, despite its simple structure, presents convergence difficulties 
for ADAPT-VQE. We evaluated both schemes using this system for three representative geometries ($r = 2.25$, $2.55$, and $2.85$~\unit{\angstrom}), shown in 
Fig.~\ref{fig:H6Chain_norm_conv}. The norm threshold is set at $10^{-4}$ (the green line in the figure). In all three geometries, the gradient norm of the 
original ADAPT-VQE stagnates well above the convergence threshold. Permanent elimination converges smoothly across all three geometries, whereas pool restoration 
introduces instabilities. Each time the pool is restored, the gradient norm surges back toward the level observed in the unmodified ADAPT-VQE before decreasing again. 
At this stage of the optimization, the ansatz has already captured the dominant correlations, so the restored operators are redundant and inflate the gradient norm 
without improving the energy. In the following sections, we therefore adopt the permanent elimination scheme for all calculations.

\begin{figure}[t!]
    \centering
    \includegraphics[width=0.8\linewidth]{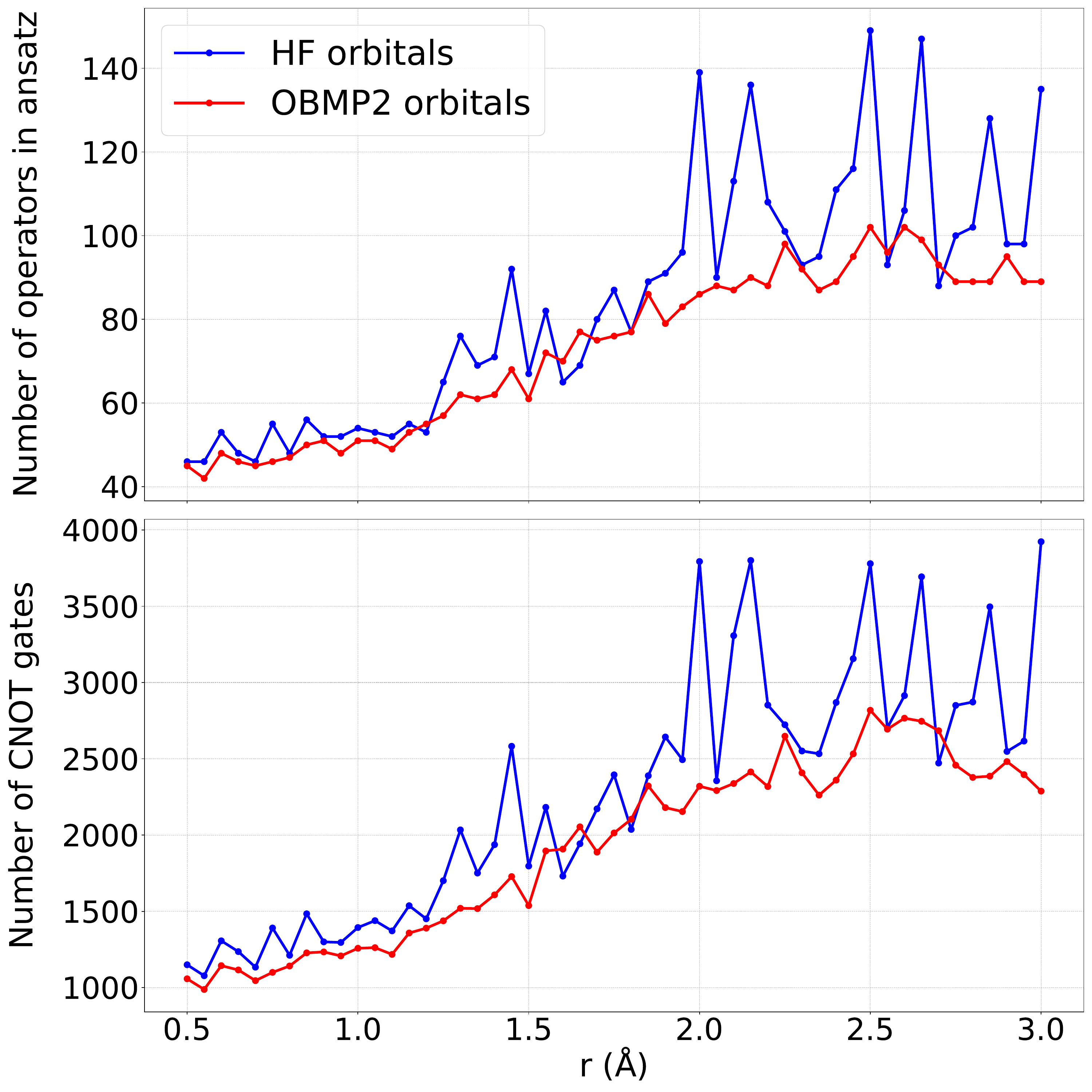}
    \caption{Number of operators and number of CNOT gates for ADAPT-VQE with the elimination scheme using different orbital bases (HF and OBMP2) at various bond lengths of $\mathrm{H_6}$ chain.}
    \label{fig:H6Chain_orb_bases}
\end{figure}

\begin{figure*}[t!]
    \centering
\includegraphics[width=0.8\linewidth]{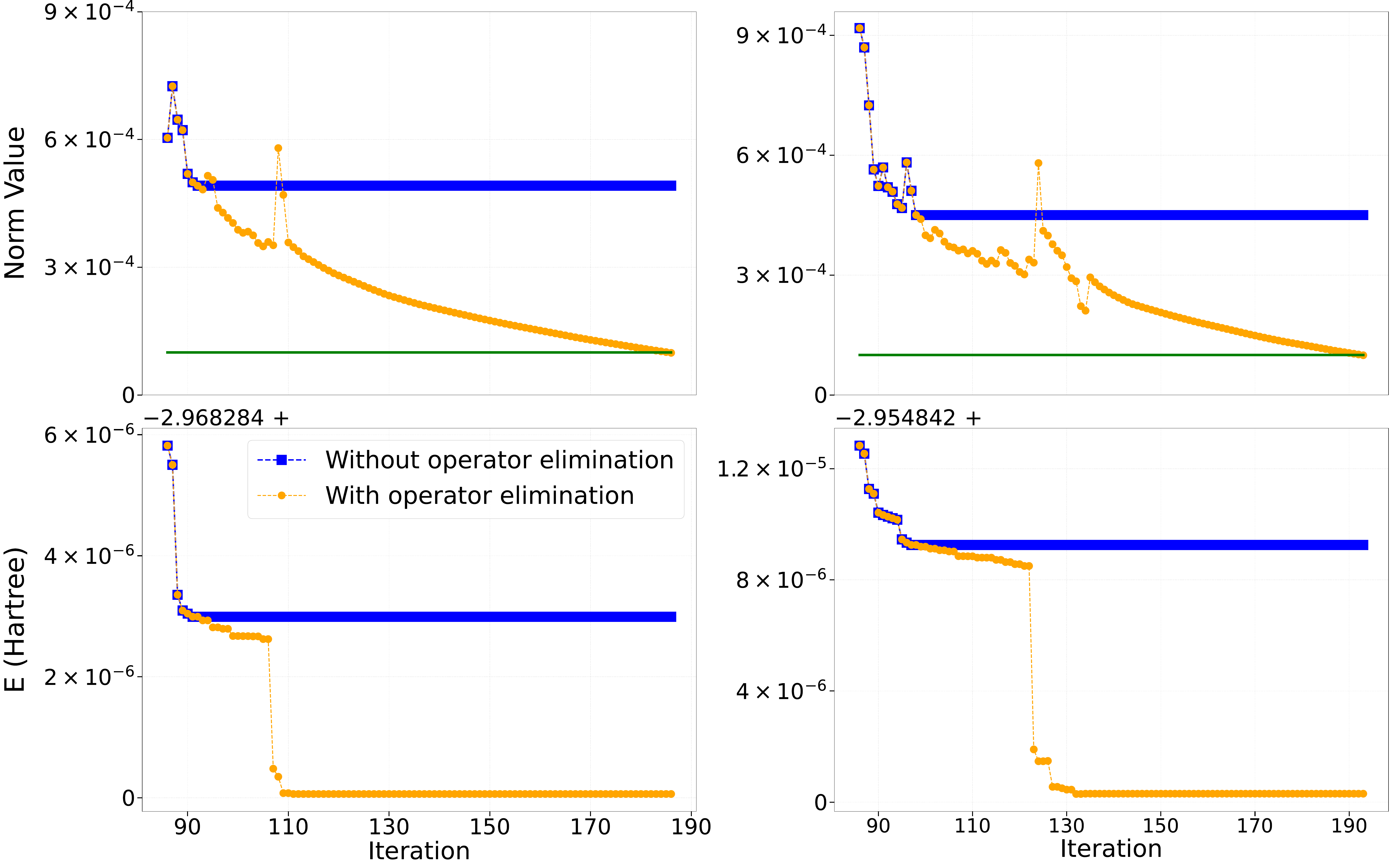}
    \caption{Convergence of the gradient norm and energy with respect to iterations 
    for $r = 2.25 \,\,\text{and}\,\, 2.45$~\unit{\angstrom} of $\mathrm{H_6}$ chain.}
    \label{fig:H6Chain_ene_conv}
\end{figure*}

\begin{figure}[t!]
    \centering
    \includegraphics[width=0.8\linewidth]{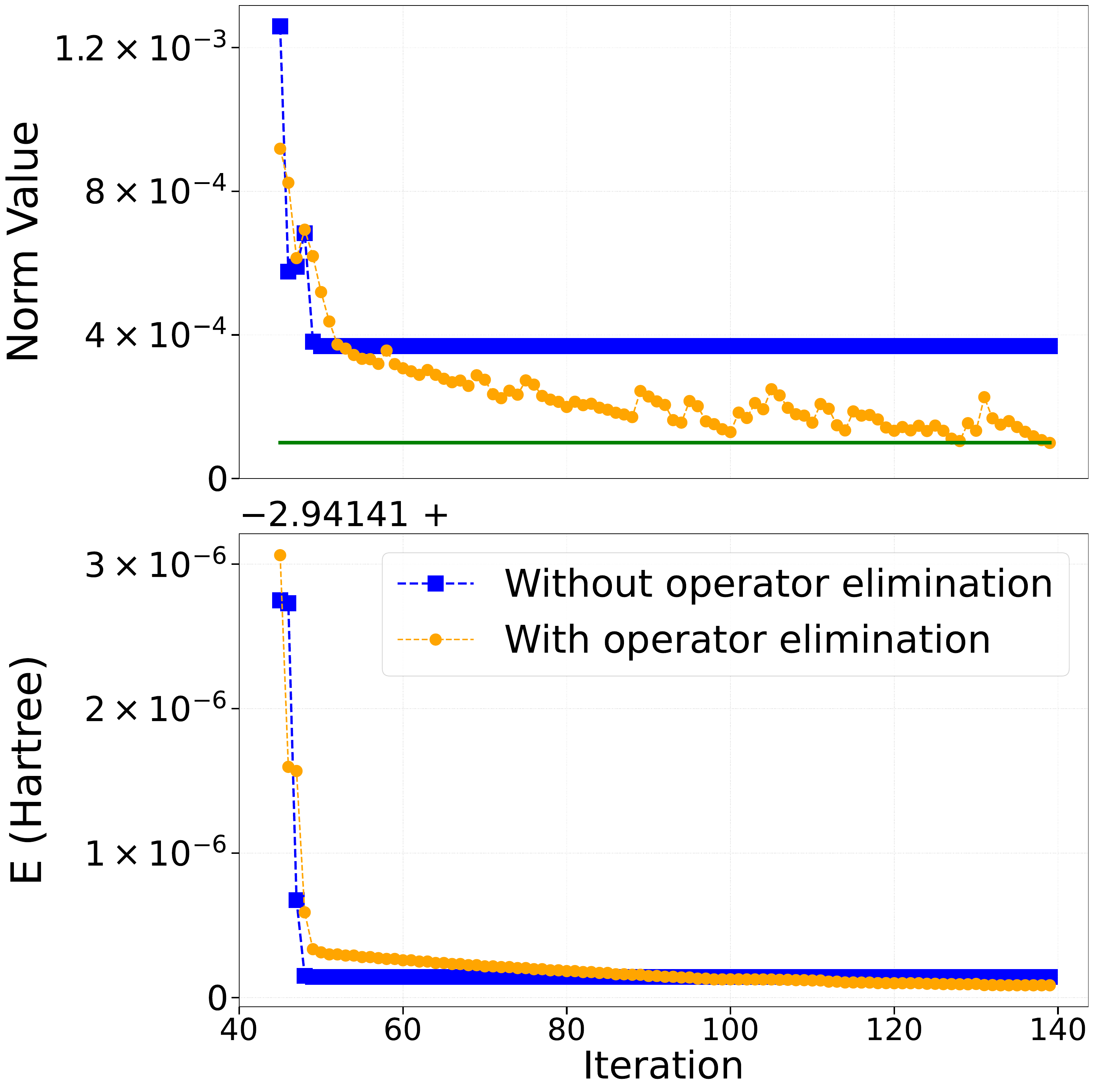}
    \caption{Convergence of the gradient norm and energy with respect to iterations 
    for $r = 3.0$~\unit{\angstrom} of $\mathrm{H_6}$ ring.}
    \label{fig:H6Ring_ene_conv}
\end{figure}

\begin{figure}[t]
    \centering
    \includegraphics[width=0.82\linewidth]{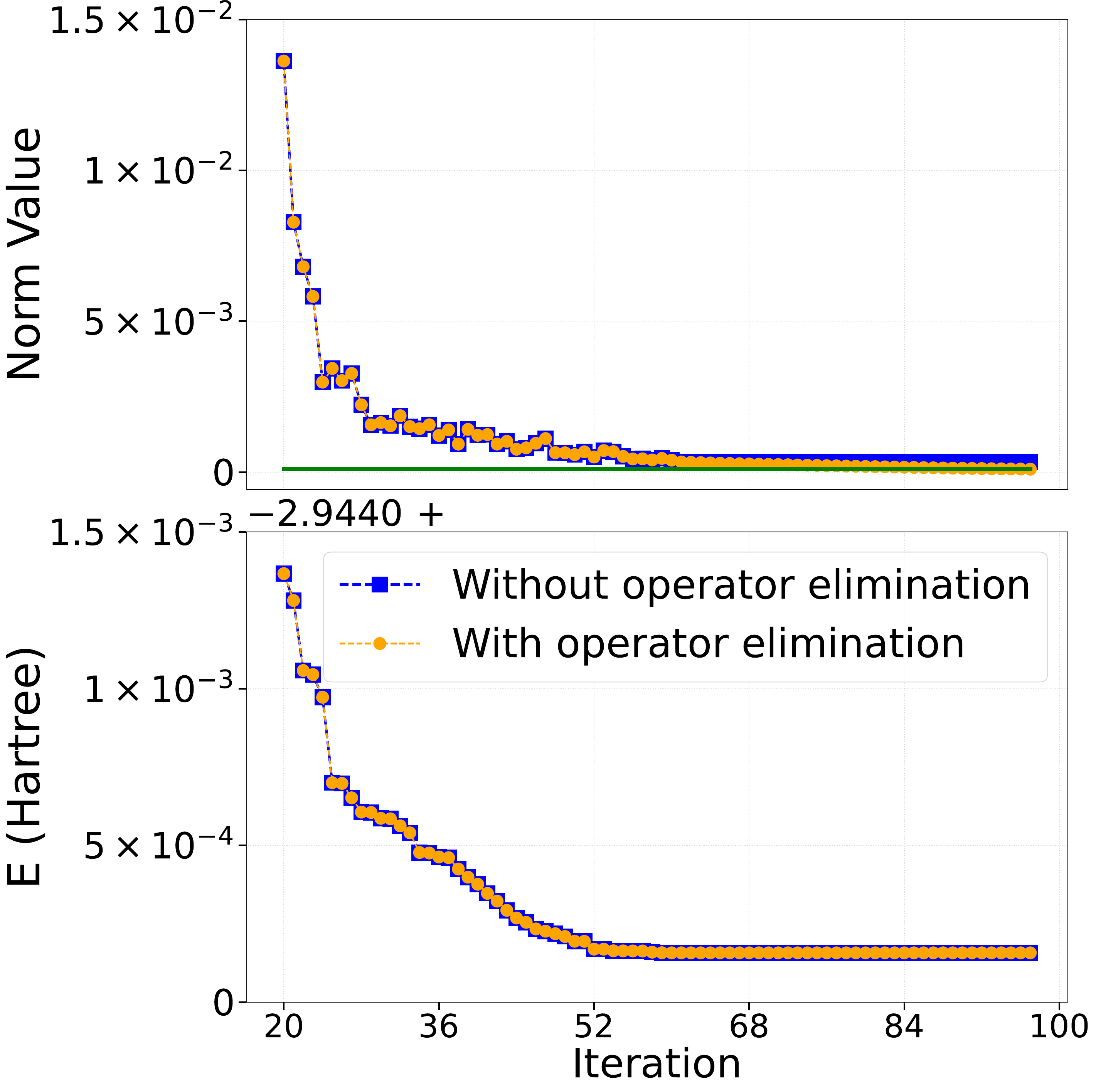}
    \caption{Convergence of the gradient norm and energy with respect to iterations 
    for $r = 2.85$~\unit{\angstrom} of $\mathrm{H_6}$ lattice.}
    \label{fig:H6Lattice_ene_conv}
\end{figure}

\begin{figure}[t]
    \centering
    \includegraphics[width=0.8\linewidth]{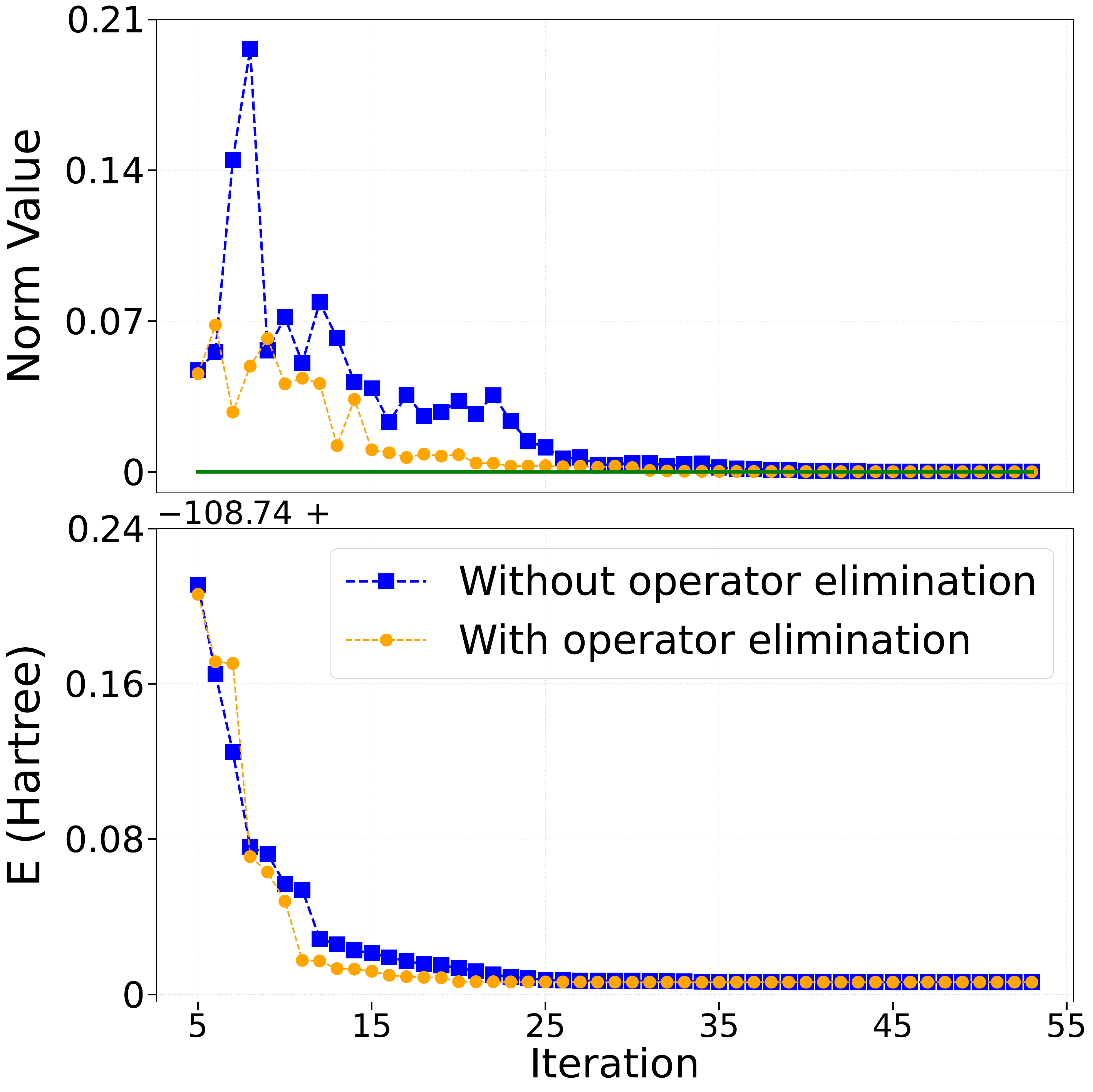}
    \caption{Convergence of the gradient norm and energy with respect to iterations for $r = 2.35$~\unit{\angstrom} in $\mathrm{N_2}$.}
    \label{fig:N2_ene_conv}
\end{figure}

\subsection{Circuit depth reduction}

We show in Fig.~\ref{fig:cnot_h6systems} the CNOT gate counts obtained from ADAPT-VQE with and without the operator elimination scheme at various bond lengths for three \ce{H6} systems. The CNOT gate counts are significantly reduced when the operator elimination scheme is employed across all systems. Without elimination (orange curves), the gate counts approach or even exceed 6000--8000 CNOT gates. In contrast, the circuit depth of ADAPT-VQE with operator elimination remains 
consistently low and relatively smooth across all bond lengths. Gate counts stay well below 4000, 3000, and 2000 for the \ce{H6} chain, ring, and lattice, respectively, representing a reduction of roughly two to three times compared to the original algorithm. The operator elimination scheme also reduces the fluctuations in the gate counts observed in the original algorithm. The slight increase in the CNOT gate count at stretched bond lengths is attributable to the strong correlation regime, which requires longer quantum circuits.

In general, a reduction in the CNOT gate count is observed for all \ce{H6} systems, implying that the direct elimination of vanishing-parameter operators is applicable 
to various molecular structures. We therefore expect that the operator elimination scheme will be practically useful for deploying ADAPT-VQE on near-term quantum devices.

\subsection{Orbital bases}

Orbital optimization is an important technique in quantum chemistry, in which the molecular orbital basis is variationally optimized alongside the wavefunction parameters. Its incorporation into VQE (OO-VQE) and ADAPT-VQE has been explored 
in recent work~\cite{sokolov2020-ooVQE,bierman2023improving,fitzpatrick2024self}. Optimizing the orbitals simultaneously with the circuit parameters can reduce quantum resource requirements and improve accuracy, particularly in strongly 
correlated regimes. However, such a procedure demands high computational costs. It is therefore highly desirable to have a pre-optimized orbital basis that can improve the performance of ADAPT-VQE. Here, we assess the performance of ADAPT-VQE with the elimination scheme using HF and OBMP2 orbitals.

Figure~\ref{fig:H6Chain_orb_bases} 
shows the operator count and CNOT gate count for ADAPT-VQE with the elimination scheme in the HF and OBMP2 orbital bases, across various bond lengths $r$(\AA) of the $\mathrm{H}_{6}$ chain. At short bond lengths ($r \lesssim 1.5$~\AA), the two bases give similar results, with operator counts between roughly 40 and 80 and CNOT gate counts between roughly 1000 and 2000.

Beyond $r \approx 1.5$~\AA, the two bases diverge noticeably. The HF orbitals (blue curve) exhibit large, irregular spikes in both the operator count and CNOT gate count across the dissociation region, reaching up to 145 operators and over 3500 CNOT gates. The OBMP2 basis, by contrast, remains considerably more compact throughout, requiring fewer operators and CNOT gates in the resulting ansatz. Overall, OBMP2 orbitals can serve as a basis for ADAPT-VQE instead of full orbital optimization, resulting in a less computationally demanding procedure.

\subsection{Energy convergence}

The natural question is whether the permanent removal of the plateau operators affects the converged energy. To address this, we compare the energy convergence of the standard ADAPT-VQE and the modified algorithm with permanent operator elimination across representative geometries of all systems studied; the results are shown in Figs.~\ref{fig:H6Chain_ene_conv}--\ref{fig:H6Lattice_ene_conv}.

The most pronounced differences are observed for the \ce{H6} chain at $r = 2.25$ and $2.45$~\unit{\angstrom} (Fig.~\ref{fig:H6Chain_ene_conv}). In both cases, the original ADAPT-VQE stalls: the gradient norm plateaus well above the convergence threshold ($10^{-4}$) and the energy remains flat over many iterations. In contrast, the modified algorithm exhibits sharp transient spikes in the gradient norm around iterations 110-130. These spikes reflect the sudden restructuring of the operator pool, and are accompanied by a small but discernible drop in energy, after which both the norm and the energy continue to decrease smoothly toward convergence. Although the energy reduction is modest, it is consistent across both geometries and is absent in the original algorithm, suggesting that the removal of small-gradient operators opens variational directions that were previously inaccessible. Overall, operator 
elimination does not compromise the converged energy and potentially lowers it further relative to the standard procedure.

\begin{figure*}[t!]
    \centering
    \includegraphics[width=0.8\linewidth]{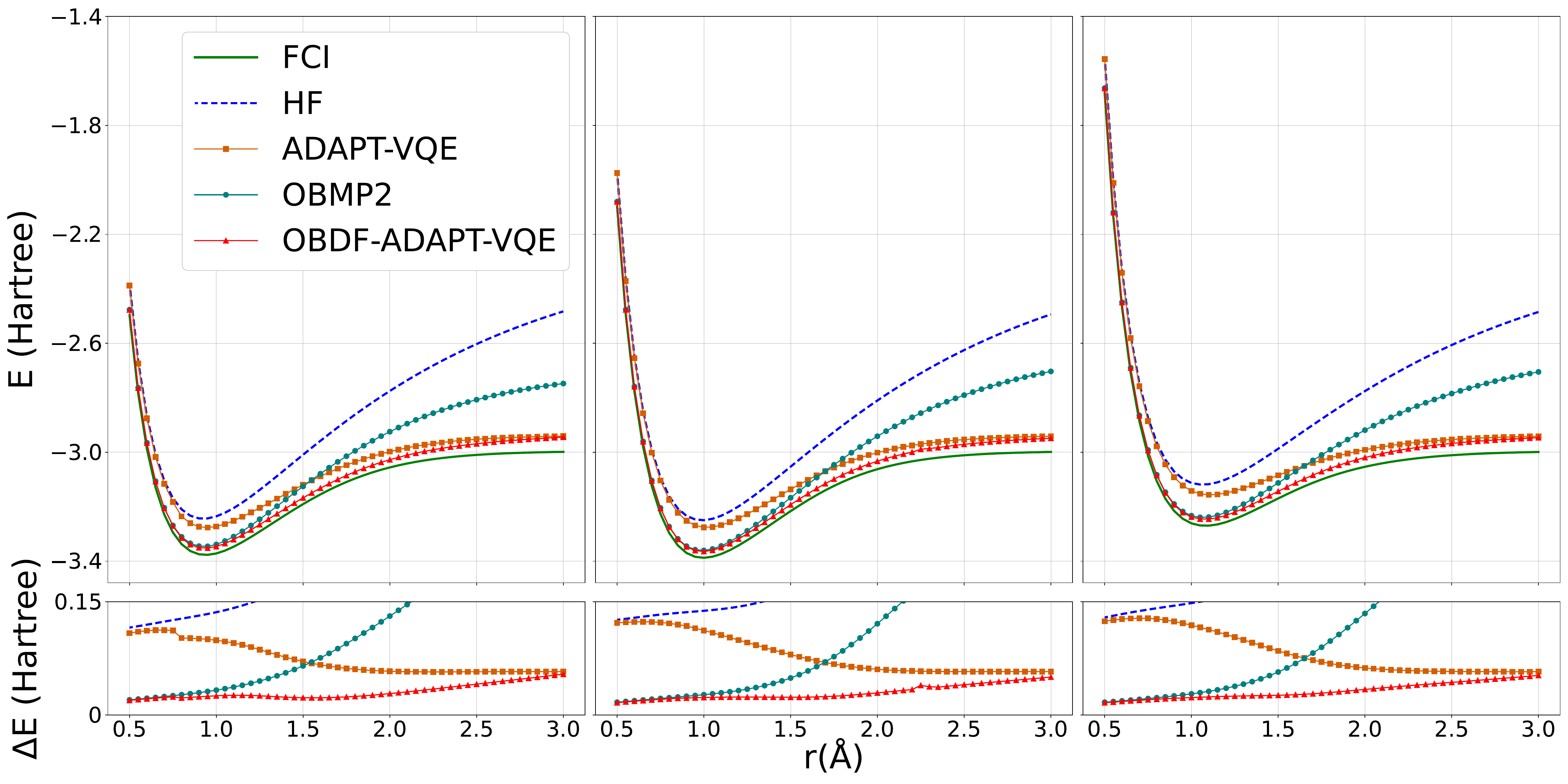}
    \caption{Potential energy surface of $\mathrm{H_6}$ systems: chain (left), ring (middle), and lattice (right). ADAPT-VQE and OBDF-ADAPT-VQE were performed for 6 spatial orbitals in active space.}
    \label{fig:PES_H6sys}
\end{figure*}

For the \ce{H6} ring at $r = 3.0$~\unit{\angstrom} 
(Fig.~\ref{fig:H6Ring_ene_conv}), the original ADAPT-VQE again fails to converge the gradient norm below the threshold, while the modified algorithm successfully drives it to convergence. 
Despite this difference in norm behavior, the two algorithms yield essentially identical converged energies. This indicates that although the original procedure becomes trapped in a plateau region, the energy at that point has already reached a value close to that of the fully converged solution. For the \ce{H6} lattice at $r = 2.85$~\unit{\angstrom} (Fig.~\ref{fig:H6Lattice_ene_conv}) and \ce{N2} at $r = 2.35$~\unit{\angstrom} (Fig.~\ref{fig:N2_ene_conv}), the original ADAPT-VQE is able to converge to the norm threshold and both schemes have almost identical behavior of convergence for both gradient norm and energy. 

In general, permanent operator elimination reproduces or even improves upon the energies obtained by the original ADAPT-VQE across systems considered in this work, with no instance in which the modification leads to a worse converged result. By reducing the effective pool size and the number of iterations required to reach convergence, the scheme yields a more efficient optimization without any energetic penalty. The permanent removal of plateau operators can therefore be regarded as a strictly non-deteriorating modification to the standard ADAPT-VQE protocol.

\subsection{ADAPT-VQE with OBDF}
Potential energy curves of H$_6$ systems are presented in Figure~\ref{fig:PES_H6sys}. We have used the active space of 6 orbitals for ADAPT-VQE and OBDF-ADAPT-VQE. Across all three geometries, ADAPT-VQE reproduces the FCI dissociation curves reasonably well at long bond lengths but exhibits large deviations near the equilibrium geometry, where dynamical correlation effects from outside the active space are important. In contrast, the OBDF-ADAPT-VQE results consistently reduce these deviations by incorporating the dynamical correlation energy from the external orbital space. It should be noted that when the distance increases beyond $2.5$~\AA\, OBDF-ADAPT-VQE demonstrates accuracy marginally superior to ADAPT-VQE, implying that the correlation captured by OBDF is still insufficient to reduce the error in stretched regime. One has to either enlarge active space or incorporate more correlation from external orbital space.

\begin{figure}[t!]
    \centering
    \includegraphics[width=0.8\linewidth]{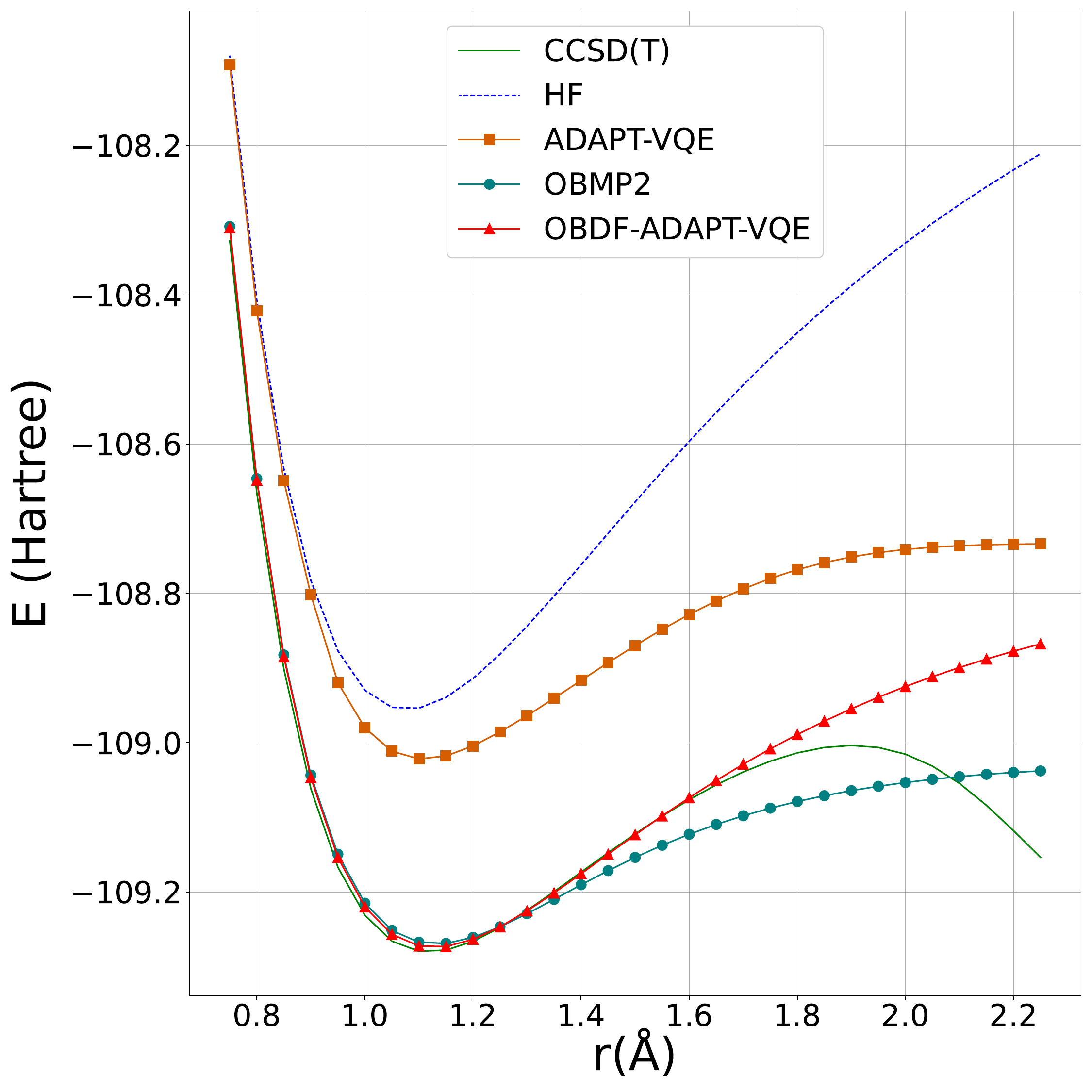}
    \caption{Potential energy surface of $\mathrm{N_2}$ molecule with respect to bond distance for 6 spatial orbitals in active space.}
    \label{fig:PES_N2Molecule}
\end{figure}

Figure~\ref{fig:PES_N2Molecule} presents the PES for the N$_2$ molecule with an active space of 6 active orbitals. The N$_2$ molecule, with its triple bond and pronounced multireference character at stretched geometries, provides a more challenging test for both methods. While CCSD(T) is a gold-standard method for near-equilibrium geometry, it breaks down rapidly in the stretched regime. OBMP2 agrees well with CCSD(T) around equilibrium and avoid the sudden divergence at long distances. However, as shown previously, OBMP2 also exhibits an unphysical barrier in the dissociation curve. ADAPT-VQE captures a correct shape of the dissociation curve but with much higher energy than CCSD(T) due to the lack of dynamical correlation. OBDF-ADAPT-VQE lowers the energy close to CCSD(T) for distances shorter than 1.6~\unit{\angstrom} and avoids the divergence displayed in CCSD(T).

In general, these results demonstrate that the integration of one-body downfolding into the ADAPT-VQE framework provides a systematic improvement in the accuracy of the computed potential energy surfaces without requiring an explicit expansion of the active orbital space. The method effectively transfers information from the inactive external orbitals into the active-space calculation through a perturbative correction, thereby improving the accuracy of the variational result. The approach is particularly beneficial when the active space is constrained by the available quantum resources, as it recovers a substantial fraction of the energy contribution from the excluded orbitals at a comparatively modest additional computational cost.

\section{Conclusion}
\label{sec:conclusion}

In conclusion, we have addressed two main drawbacks of the standard ADAPT-VQE algorithm: redundant operator and the neglect of dynamical correlation from outside the active space. To mitigate the former, we introduced a simple and direct operator elimination strategy that dynamically removes operators with vanishing parameters from the pool. Benchmarks on the linear H$_6$ chain, the H$_6$ lattice, the H$_6$ ring, and the N$_2$ molecule in various geometries show that this strategy enables optimization to converge smoothly without exhausting iterations on redundant operators that do not  lower total energy. To address the second limitation, we combined ADAPT-VQE with OBDF, in which the bare molecular Hamiltonian in active space is replaced by a correlated effective Hamiltonian that includes the dynamical correlation effect outside the active space. Across all benchmark systems, ADAPT-VQE with OBDF yields energies closer to the FCI reference than conventional ADAPT-VQE within the same active space. In general, the combination of the two strategies proposed in this work is promising toward more practical variational quantum simulations of molecular systems.

\section*{Acknowledgments}
This research is funded by the National Foundation for Science and Technology Development (NAFOSTED) Grant Number 103.01-2024.06. The authors used AI-assisted tools only for manuscript language editing.

\bibliography{main}
 
\end{document}